\documentclass[doublecol]{epl2} 
\newcommand{\bea}{\begin{eqnarray}}    
\newcommand{\eea}{\end{eqnarray}}      
\newcommand{\be}{\begin{equation}}
\newcommand{\ee}{\end{equation}}
\newcommand{\bef}{\begin{figure}}
\newcommand{\eef}{\end{figure}}

\def\spose#1{\hbox to 0pt{#1\hss}}
\def\ltapprox{\mathrel{\spose{\lower 3pt\hbox{$\mathchar"218$}}
\raise 2.0pt\hbox{$\mathchar"13C$}}}
\def\gtapprox{\mathrel{\spose{\lower 3pt\hbox{$\mathchar"218$}}
\raise 2.0pt\hbox{$\mathchar"13E$}}}
\def\inapprox{\mathrel{\spose{\lower 3pt\hbox{$\mathchar"218$}}
\raise 2.0pt\hbox{$\mathchar"232$}}}

\usepackage{color}

\begin{document}

\title{Very large scale correlations in the galaxy distribution} 
\shorttitle{Large scale galaxy correlations}

\author{Francesco Sylos Labini\inst{1,2} } 
\shortauthor{Sylos Labini}
\institute{ \inst{1} Museo
  Storico della Fisica e Centro Studi e Ricerche Enrico Fermi, -
  Piazzale del Viminale 1, 00184 Rome, Italy \\ \inst{2} Istituto dei
  Sistemi Complessi CNR, - Via dei Taurini 19, 00185 Rome, Italy }

\abstract{We characterize galaxy correlations in the Sloan Digital Sky
  Survey by measuring {  several moments of} galaxy counts in
  spheres.  We {  firstly} find that the {  average counts} grows
  as a power-law function of the distance with an exponent $ D= 2.1
  \pm 0.05 $ for $r \in [0.5,20]$ Mpc/h and $D = 2.8 \pm 0.05 $ for
  $r\in [30,150]$ Mpc/h. {  In order to estimate the systematic
    errors in these measurements we consider } the counts variance
      {  finding that it} shows systematic finite size effects which
      depend on the samples sizes.  We clarify, by making specific
      tests, that these are due to galaxy long-range correlations
      extending up to the largest scales of the sample. The analysis
      of mock galaxy catalogs, generated from cosmological N-body
      simulations of the standard LCDM model, shows that for $ r<20$
      Mpc/h the counts exponent is $D\approx 2.0$, weakly dependent on
      galaxy luminosity, while $D=3$ at larger scales. In addition,
      contrary to the case of the observed galaxy samples, no
      systematic finite size effects in the counts variance are found
      at large scales, a result that agrees with the absence of large
      scale ($r\approx 100$ Mpc/h) correlations in the mock catalogs.
      We thus conclude that the observed galaxy distribution is
      characterized by correlations, fluctuations and hence
      structures, which are larger, both in amplitude and in spatial
      extension, than those predicted by the standard model LCDM of galaxy
      formation.  }

\maketitle

\section{Introduction} 

The quantification of large scales galaxy correlations and
fluctuations is a central problem in cosmology. The advent of massive
redshift surveys has made possible precise measurements of the galaxy
two-point correlation function on ten Mpc scale, where power law
correlations (in redshift space) have been well established
\cite{zehavi,Hawkins,dr4_paper}. On larger scales the situation is
less clear as large density fluctuations on 150 Mpc/h scales were
observed
\cite{shanks1,shanks2,shanks3,einasto,2df_epl,2df_aea,saslaw,einasto11}:
these are not obviously compatible with the absence of strong
clustering on those scales as predicted by standard models of galaxy
formation \cite{pee93,glass}.  In addition, in a deep sample of very bright
galaxies it was observed \cite{martinez_bao,bao_aea,kazin}, on 200
Mpc/h scales, an unexpected strong signal with respect to the
predictions of the standard model; recently an excess of clustering
was also found by \cite{shanks4,lahav11}.

{  In order to characterize} the large scale galaxy correlations it
was measured, in the Sloan Digital Sky Survey (SDSS) \cite{york,dr6},
the average density $\overline{n(r)}$: this was shown to present a
scaling behavior as $\overline{n(r)} \sim r^{-1}$ for $r<20$ Mpc/h
\cite{hogg,dr4_paper}.  However, while some authors \cite{hogg}
noticed that around 70 Mpc/h it occurs a transition toward uniformity
(i.e. $\overline{n(r)} \sim const.$), others concluded that
$\overline{n(r)} \sim r^{-0.2}$ for $20$ Mpc/h $<r<80$ Mpc/h
\cite{tibor}. Furthermore this latter behavior was shown to be
associated to a scaling of the variance slower than for a uniform
distribution and to a probability density function (PDF) of galaxy
counts in spheres well fitted by a Gumbel distribution (i.e.,
different from a simple Gaussian PDF) \cite{tibor}.  It was then 
found that the PDF, in different spatial volumes, present systematic
differences at large enough scales which, by making specific tests,
were interpreted as due to large scale structures
\cite{sdss_epl,sdss_aea}.  However, the considered tests used
non-overlapping sub-samples covering different redshift ranges, so
that a physical or observational scale dependent systematic effect 
can, at least partially, affect the {  observed behaviors}. 

For instance, it was noticed by that number density of bright galaxies
increases by a factor $\approx $ 3 as redshift increases from $z = 0$
to $z = 0.3$ \cite{loveday}, and to explain these observations a
significant evolution in the luminosity and/or number density of
galaxies at redshifts $z < 0.3$ was proposed. In this context
evolution can be parameterized as a redshift dependent correction
\cite{blanton2003}.  However, by performing several tests to determine
the possible effect of evolution on the SDSS data, it was concluded
\cite{sdss_aea} that while the evolution may change {  the galaxy
  redshift counts behavior as a function of redshift, but} it is
unlikely that it can produce the large amplitude fluctuations of large
spatial extension.

In this letter, by considering SDSS equal volume samples
covering the same distance scales, we are able to disentangle
finite-size effects due to large scale correlations from redshift
dependent systematic effects. This allows us to confirm the results by
\cite{sdss_aea,tibor}, to extend them to larger scales, i.e. $r\approx
150$ Mpc/h, and to clarify several other properties of the large scale
galaxy distribution.

\section{The Data} 


We have constructed several sub-samples of the main-galaxy (MG) sample
of the spectroscopic catalog SDSS-DR7 \cite{strauss,dr7}. The
selection criteria used to construct the volume limited
(VL) 
\footnote{ {  The prescription used to define a volume limited
    sample takes into account the luminosity selection effects that
    necessarily affect a survey, i.e. that intrinsically faint
    galaxies are observed only if they are located close to the
    observer while intrinsically bright galaxies are observed both if
    they are nearby to and faraway from the observer.}}
 samples are (see \cite{sdss_aea}
  for more details): we selected galaxies from the MG sample with
  redshift $z\in [10^{-4},0.3]$ with redshift confidence $z_{conf} \ge
  0.35$, without significant redshift determination errors and with
  apparent magnitude $m_r < 17.77$. In order to avoid the irregular
  edges of the survey boundaries, and to consider a simple sky area,
  we have constructed a sample ($R_1$) limited, in the internal
  angular coordinates of the survey \cite{sdss_aea}, by $\eta \in
  [-33,5^\circ,36.0^\circ]$ and $\lambda \in [-48^\circ,51.5^\circ]$.
We also consider in what follows two half samples of equal volume are
limited by ($R_2$) $\lambda \in [-48^\circ,0]$ and ($R_3$) $\lambda
\in [0^\circ,51.5^\circ]$. 

{  The survey is known to have a small angular incompleteness:
  indeed, in average, about the $5\%$ of the target galaxies have not
  measured redshift \cite{strauss}.  In general, there is not a free
  of a-priori assumptions procedure, to correct for such an
  incompleteness. Indeed, given that the detailed information of the
  real galaxy distribution is unknown one has to make some assumptions
  on the statistical properties of such a distribution (see
  e.g. \cite{zehavi}). As we do not want to use such ad-hoc
  assumptions, as we employ statistical methods that aim to tests some
  of the most common assumptions in the analysis of galaxy
  correlations \cite{book}, we have adopted the following
  strategy. From the one hand we have considered an angular region
  which does not include the survey edges where completeness varies
  mostly. From the other hand we have done a few tests to control the
  effect of completeness on the correlation analysis, and in
  particular, as we discuss below, we have focused to test the {\it
    stability} the results.  Note that the incompleteness due to fiber
  collisions can be neglected in measurements of large-scale (i.e.,
  $r>10$ Mpc/h) galaxy correlations as this effects is only relevant
  very small scales (see \cite{sdss_aea} and references therein).}

In order to construct VL samples (see
Tab.\ref{tbl_VLSamplesProperties1}) we computed the metric distances
$R$ using the standard cosmological parameters, i.e., $\Omega_M=0.3$
and $\Omega_\Lambda=0.7$.  Absolute magnitudes $M$ are computed using
Petrosian apparent magnitudes in the $m_r$ filter corrected for
Galactic absorption and applying standard K-corrections \cite{vagc}.
\begin{table}
\begin{center}
\begin{tabular}{|c|c|c|c|c|c|}
  \hline
  VL sample & $R_{min}$ & $R_{max}$ & $M_{min}$ 
& $M_{max}$ & $N_p$ \\
  \hline
    VL1    & 50  & 200 & -18.9 & -21.1   & 73811  \\
    VL2    & 125 & 400 & -20.5 & -22.2   &  129975  \\
    VL3    & 200 & 600 & -21.6 & -22.8   & 51698 \\
    \hline
\end{tabular}
\end{center}
\caption{Main properties of the SDSS VL samples : 
$R_{min}$, $R_{max}$ (in Mpc/h) are the chosen limits for the metric
distance; ${M_{min}, \,M_{max}}$ define the interval for the absolute
magnitude in each sample and $N_p$ is the number of galaxies in the sample.}
\label{tbl_VLSamplesProperties1}
\end{table}

\section{Statistical methods}

We consider the number of galaxies $N_i(r)$ within radius $r$ around
the $i^{th}$ galaxy: this quantity differs for each galaxy and hence
we consider it as a random variable characterizing its statistical
properties.  The average over an ensemble of realizations, $\langle
N(r) \rangle$, can be estimated by the volume average (assuming
ergodicity)\footnote{The symbols $\langle ... \rangle$
  ($\overline{...}$) stands for the ensemble (volume) average
  performed with the condition that the sphere center coincides with a
  point of the distribution, i.e. it is a conditional average
  \cite{book}.}
\begin{equation}
\label{eq:condden}
\overline{N(r)}= \frac{1}{M(r)} \sum_{i=1}^{M(r)} N_i(r) \;. 
\end{equation}
Note that the number of points contributing to the average
(Eq.\ref{eq:condden}), $M(r)$, depends on the scale $r$ as an effect
of the requirement that the spheres must be fully included in the
sample's boundaries \cite{sdss_aea}.
To estimate the typical fluctuations of the random variable $N_i(r)$
we may consider the conditional variance $\Sigma^2(r) = \langle N^2(r)
\rangle - \langle N(r) \rangle^2 \;.$ In general \cite{book}, this can
be written as the sum of two contributions: $\Sigma^2(r) =
\Sigma_i^2(r) + \Sigma_p^2(r)$, where $\Sigma_i(r)$ is the intrinsic
part,  due to correlations, and $\Sigma_p^2(r) = \langle
N(r) \rangle$, due to Poisson noise.  The normalized variance is
defined as $\sigma^2(r) = {\Sigma^2(r)}{\langle N(r) \rangle^{-2}} =
\sigma_i^2(r) + \sigma_p^2(r) \;,$ where its intrinsic part can be
estimated by 
\begin{equation}
  \label{eq:condvar}
 \overline{\sigma_i^2(r)}  = 
\frac{1}{ \overline{N(r)}^2}\times 
\frac{
  \sum_{i=1}^{M(r)} (N_i - {\overline{N(r)}})^2 }{M(r)}
-\frac{1}{\overline{N(r)}}  
\;.
\end{equation}

\section{Results}

\begin{figure}
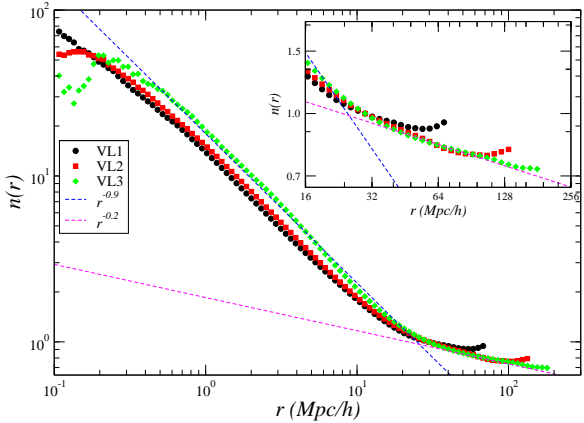

\onefigure[scale=0.32]{fig1.eps}
\caption {Galaxy average density normalized to its value at 30
  Mpc/h, for the different SDSS samples. The exponent $\gamma$
  reported in the labels corresponds to the best fit in the
  range [0.5,20] Mpc/h.}
\label{Gamma_all}
\end{figure}

We have measured the conditional average density, $\overline{n(r)} =
\overline{N(r)}/{V(r)}$ where $V(r) = 4/3 \pi r^3$, in the different
SDSS VL samples (see Fig.\ref{Gamma_all}).  We find that: (i) at small
scales, $r\in [0.5,20]$ Mpc/h $ \overline{n(r)} \propto r^{-\gamma}$
with $\gamma = 0.88 \pm 0.05$ (i.e. $\overline{N(r)} \propto r^D$,
with $D=3-\gamma=2.12\pm 0.05$). The error of the exponent has been
derived by measuring the scattering in the different samples.

(ii) At larger scales, i.e. $r>30$ Mpc/h and up to $\approx 150$ Mpc/h
in the deepest sample, the exponent is $\gamma=0.2 \pm 0.05$
(i.e. $D=2.8\pm0.05$).  The amplitude of the conditional average
density in the different VL samples is in principle fixed by a
luminosity factor that depends on the absolute luminosity of the
galaxies therein included \cite{sdss_aea}. However, in a finite
sample, {  as long as correlations extend on scales of the order of
  the sample size}, the amplitude of $\overline{n(r)}$ also depends on
the particular structures present in that specific volume. { 
  Therefore} the {  galaxy counts} normalization in samples of
different sizes, covering different space volumes and including
galaxies of different luminosity depends on systematic effects which
become negligible only for very large sample sizes.  For this reason
in Fig.\ref{Gamma_all}, by choosing an arbitrary normalization, we
have plotted $\overline{n(r)}/\overline{n(r_*)}$ where $r_*=30$ Mpc/h.
On the other hand, note that the {\it exponent } of $\overline{n(r)}$
does not show variations larger than the estimated error bar (see
below) in the different samples.


The main result from this analysis is that the {\it slopes} of the
galaxy average density, both at small and large scales, show a very
good agreement in the different samples.  When the radius $r$
approaches the size of the largest sphere included in each sample
there are systematic effects, as shown by the fact that the large
scale tail of $\overline{n(r)}$ (where  $M(r)$ in Eq.\ref{eq:condden}
 becomes rapidly very
small) grows at different scales in the different samples.

\begin{figure}
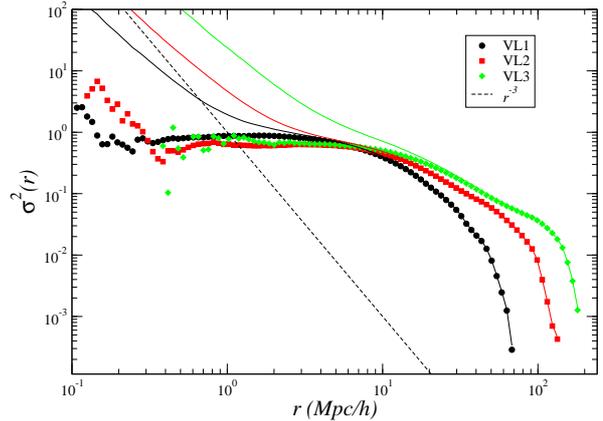

\onefigure[scale=0.32]{fig2.eps}
\caption {Normalized conditional variance (Eq.\ref{eq:condvar}) in the
  different samples: circles represent the intrinsic variance
  $\overline{\sigma_i^2(r)}$, while the solid lines the total variance
  $\overline{\sigma^2(r)}$ (see Eq.\ref{eq:condvar}). The dashed line
  has a slope $1/r^3$ as for pure Poisson noise, a behavior that
  describes only the very small scales.}
\label{Sigma_all}
\end{figure}

{  In order to} quantify fluctuations affecting the large scale
(i.e. $r> 10$ Mpc/h) behavior of $\overline{n(r)}$ it is necessary to
estimate statistical errors. For instance, one may simply compute them
as $\sqrt{\Sigma^2(r)V^{-2}(r) M^{-1}(r))}$: however, in this way one
underestimates the true errors because large scale correlations can
break the Central Limit Theorem and thus the different determinations
are not independent \cite{book} (note that such error bars would not
be even visible in the plot in Fig.\ref{Gamma_all}). A {  possible
  evaluation} of the scattering of $\overline{n(r)}$ can be { 
  performed} by measuring sample-to-sample variations \cite{bao_aea}
. To this aim, we calculate thus Eq.\ref{eq:condvar} in the different
samples (see Fig.\ref{Sigma_all}) finding that at small scales results
are very similar while at large scales a clear difference is detected.

\begin{figure}
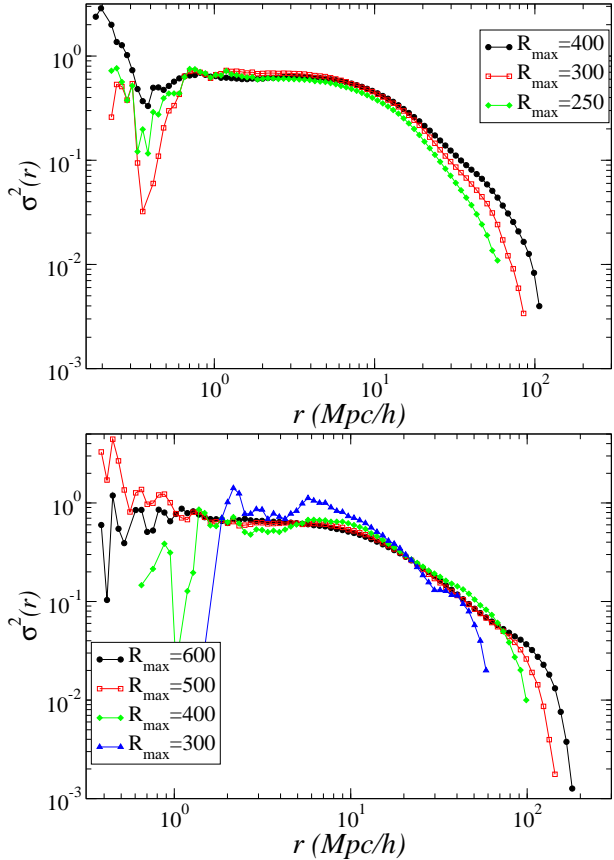

\onefigure[scale=0.32]{fig3a.eps}
\onefigure[scale=0.32]{fig3b.eps}
\caption {Normalized conditional variance in
  subsamples of VL2 (upper panel) respectively with $R_{max}=250,300,400$
  Mpc/h and VL3 (bottom panel) with $R_{max}=300,400,500,600$ Mpc/h. }
\label{cv_fs} 
\end{figure}

Interestingly the differences in $\overline{\sigma_i^2(r)}$ occur at a
scales that grows with sample's sizes. The determination in the
smallest sample, i.e. VL1, shows a marked large scale difference with
respect to those in VL2 and VL3, which instead present a more similar
behavior everywhere but the very large scale tail ({  i.e. $r\approx 
  100$ Mpc/h}).  Therefore a possible explanation of this large scale
behavior is that it is due to a finite size effect, i.e. that the
scale at which the abrupt decay of $\overline{\sigma_i^2(r)}$ occurs
in the different samples, depends on the their sizes.
To show that this is the case we have considered the behavior of
$\overline{\sigma_i^2(r)}$ in sub-volumes of the VL2 and VL3 samples
limited to a depth 
(i.e., $R_{max}$) smaller
than the ones reported in
Tab.\ref{tbl_VLSamplesProperties1}.  In this way we are able to
clearly single out the finite size effect (see Fig.\ref{cv_fs}).

\begin{figure}
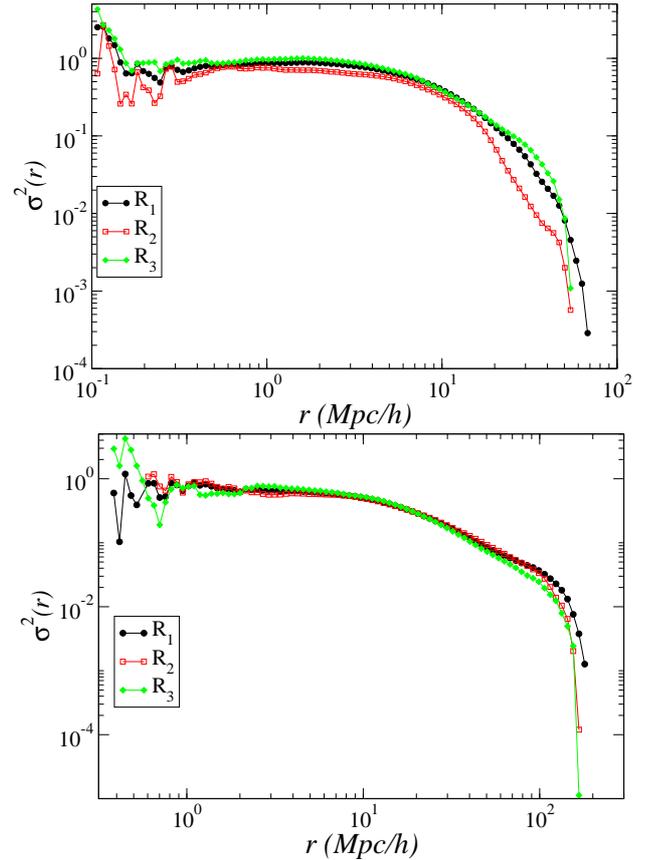

\onefigure[scale=0.32]{fig3c.eps}
\onefigure[scale=0.32]{fig3d.eps}
\caption {Normalized conditional variance in the three
  regions of VL1 (upper panel) and VL3 (bottom panel). }
\label{cv_fs2} 
\end{figure}

We performed another test by cutting the sample into two subsamples of
equal volume (i.e., limited in angles by the regions $R_2$ and $R_3$)
and determining $\overline{n(r)}$ and $\overline{\sigma_i^2(r)}$ in
each of them (see Fig.\ref{cv_fs2}). In the case of VL3 the
determinations of both the average density and the variance in two
subsamples agree better than for VL1: this corroborates the result,
obtained also with the previous tests, that systematic finite size
effects become weaker in samples with larger volumes
\cite{sdss_aea,tibor,copernican}. {  This is in agreement with the
  larger dimension observed at large scale, which also implies less
  wild fluctuations \cite{book}}.

From these behaviors we may conclude that both 
$\overline{n(r)}$ and $\overline{\sigma^2(r)}$ converge
to a well-defined value only for scales $r<r_s$ smaller than the
radius of the largest sphere included in the sample.  We estimate
that for  
VL1  $r_s\approx 50$ Mpc/h, for VL2 $r_s\approx 100$ Mpc/h and
for VL3 $r_s\approx 150$ Mpc/h. Beyond these scales the behaviors in
the two half samples systematically differ: this is in agreement with
the results by
\cite{sdss_aea,sdss_epl,2df_aea,2df_epl,tibor,copernican} where the
whole behavior of the PDF of $N_i(r)$ was considered.

{  As mentioned above it is interesting to consider the effect of
  the small (i.e. $\sim 5\%$) angular incompleteness on the
  measurements of galaxy correlations. For instance, there are some
  small sky regions where, due to the presence of bright stars,
  galaxies have not been observed. Each bright star corresponds to a
  circular hole where galaxies cannot be observed.  The radius of the
  circle around each bright star depends on the star apparent
  magnitude \cite{vagc}. An upper limit to such a size is $\theta = 3$
  arsec, corresponding to very bright stars \cite{vagc}.  A simple way
  to test for effect of such holes is the following.

 We distribute randomly holes of size $\theta$ with a surface density
 of $50$ per square degree (roughly corresponding the surface density
 of bright stars \cite{hog}).  Galaxies which are placed inside one of
 the holes are cut from the resulting sample. In such a way we
 artificially have taken away from the sample, in a correlated way at
 small angles, about $\sim 10\%$ of the galaxies (i.e. about the
 double than the angular incompleteness of the catalog). The question
 is whether the large scales (i.e., $r>$1 Mpc/h) correlation
 properties are affected by such an incompleteness.  Results are shown
 in Fig.\ref{Gamma_holes}: one may note that, a part an obvious $10\%$
 shift in the amplitude of the average conditional density, no
 scale-dependent changes are manifested. Thus we can confidently
 conclude that incompleteness effects do not play a major role in the
 results of the correlation analysis. We refer to a forthcoming work
 for a more complete series of tests on the angular incompleteness of
 this survey.
\begin{figure}
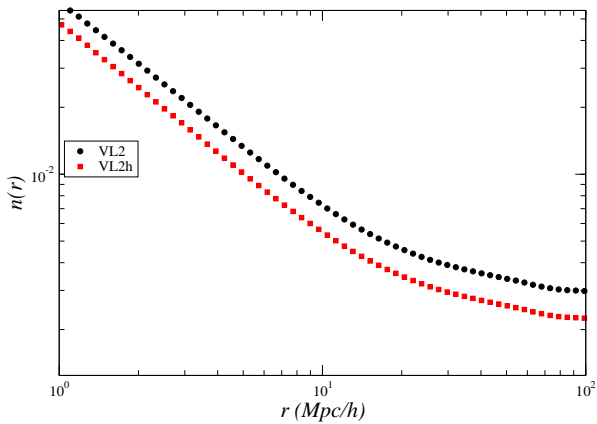

\onefigure[scale=0.32]{fig3bis.eps}
\caption {  Conditional average density for the full sample VL2 and for
  a modified version of it (VL2h) where $\sim 10\%$ of the galaxies
  have been cut in a correlated way (see text): a part an obvious
  $10\%$ shift in the amplitude of the average conditional density,
  not scale-dependent changes are manifested.}
\label{Gamma_holes}
\end{figure}
}

\section{Mock galaxy catalogs} 

\begin{figure}
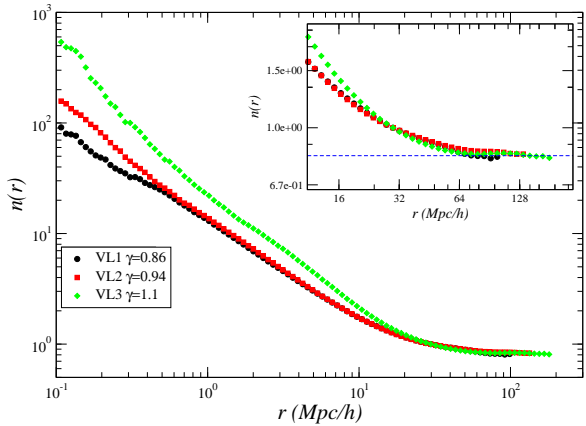

\onefigure[scale=0.32]{fig4.eps}
\caption {Conditional average density, normalized to its value at 30
  Mpc/h, for the different samples of the mock galaxy catalog. The
  exponent $\gamma$ reported in the labels is the best fit in the
  range [0.5,20] Mpc/h}
\label{Gamma_all_mock}
\end{figure}

\begin{figure}
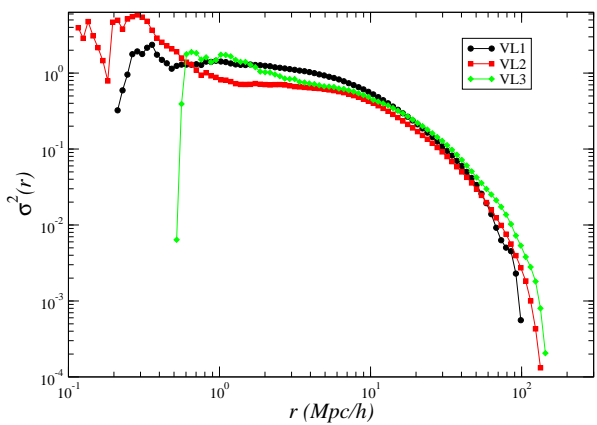

\onefigure[scale=0.32]{fig5.eps}
\caption {Normalized conditional variance for the different samples of
  the mock galaxy catalog. No finite size effects are present in this
  case.}
\label{Sigma_all_mock}
\end{figure}

We have repeated the same analysis in some mock galaxy samples.  These
are constructed from cosmological N-body simulations of the standard
LCDM model \cite{springel,croton}, by applying the same cuts in
absolute magnitude and distance as those reported in
Tab.\ref{tbl_VLSamplesProperties1} and by computing the redshift
positions (i.e., simply applying the corrections to the redshift due
to the peculiar velocities along the line of sight).  The conditional
average density (see Fig.\ref{Gamma_all_mock}) shows (i) a slope
$\gamma \approx 1$ for $r\in[0.5,20]$Mpc/h that is weakly dependent on
the average luminosity of the galaxies, and (ii) a well defined
crossover to uniformity, i.e. $\gamma=0$, at $\sim 30$ Mpc/h. Both
features are different from the real SDSS samples.
Correspondingly to the crossover toward uniformity (absence 
of large scales strong correlations), 
the conditional variance (Fig.\ref{Sigma_all_mock}) does
not show a finite size dependence, contrary to what 
occurs for the real samples (Fig.\ref{Sigma_all}).
In this case, the stability of the normalized
variance in the different samples is shown by diving the samples into
two  parts and by comparing the behaviors: 
because of the lack of large scale correlations, no systematic
differences are found between the two subsamples.

{  Note that a given mock galaxy sample is constructed from the
  underlying dark matter particles distribution, by assuming a certain
  prescription to identify galaxies.  Such a prescription is generally
  based on a physical mechanism which links the local density of dark
  matter particles to the probability to form a mock galaxy
  \cite{croton}.  In principle, one can introduce different physically
  motived prescriptions to relate dark to visible matter distributions
  in cosmological simulations. The question is then whether a
  different prescription can substantially change the resulting mock
  galaxies correlation properties, possibly giving rise to a better
  agreement with observations.  We do not explore this question here
  and we cannot exclude that a better agreement could be obtained with
  a different bias prescription. However we note that, given that
  large-scale correlations are not present in the underlying dark
  matter distribution, as this is what theoretical models predicts on
  scales of the order of $\sim 100$ Mpc/h \cite{glass,cdm_theo}, the
  only way to introduce such correlations is via a bias mechanism:
  biasing should correspond to a large scale correlated
  selection. This implies the biasing mechanism to be non-local,
  i.e. a completely different one from what is generally considered to
  be a physically plausible prescription.}

\section{Discussion} 

In summary, we have analyzed the scaling properties of the galaxy
distribution in the SDSS-DR7 samples.  We found that at small scales
$r\in[0.5,20]$ Mpc/h the galaxy (conditional) average density decays
as a power-law function of distance, $\overline{n(r)} \propto r^
{-\gamma} $ with an exponent $\gamma\approx 0.9$ while at large
scales, i.e.  $r \in [30,150]$ Mpc/h, it decays slower: $\gamma
\approx 0.2$.  The analysis of the variance (Eq.\ref{eq:condvar})
allows to clearly single out a finite size effect, that can be simply
understood as due galaxy correlations extending up to the largest
scales of the considered samples.  These behaviors differ from those
found in mock galaxy samples where: (i) at small scales the
exponent, $\gamma \approx 1$, slightly depends on galaxy luminosity,
and (ii) at large scales, i.e. $r>30$ Mpc/h, a well defined crossover
to uniformity (i.e. $\gamma=0$) is found. In addition,
no finite size effects
are detected in the large scale behavior of the normalized counts
variance.

The detection of very large scale galaxy correlations and of
finite-size effects, allow us to conclude that the observed galaxy
distribution is more correlated, fluctuating and thus clustered, than
the one predicted by the standard LCDM model of galaxy formation
through cosmological N-body simulations \cite{springel,croton}. These
findings are in agreement with the results obtained in the same
samples, although at smaller scales $r<100$ Mpc/h, through different
tests \cite{sdss_aea,sdss_epl,tibor,copernican,saslaw}, with the
results obtained in the 2 degree field redshift survey \cite
{shanks1,shanks2,shanks3,2df_epl,2df_aea,einasto} and other surveys
\cite {martinez_bao,bao_aea,kazin,shanks4,lahav11}. Note that the
results obtained by \cite{hogg} are also compatible with a slow decay
of $\overline{n(r)}$ at large scales.

Because of the large-scale scaling of the galaxy average density we
conclude that any statistical quantity normalized to the estimation of
the sample average density (e.g., the standard two-point correlation
function) is biased by finite size effects \cite{book}. This implies
that the volume of current galaxy samples is not yet large enough to
measure the standard two-point correlation function at $\sim 100$
Mpc/h.  

From a theoretical point of view, the main challenge of our results
concerns the way in which the large scale universe is modeled. Indeed
the deviation from a pure statistically homogeneous and isotropic, and
spatially uniform density field \cite{copernican} imply the
consideration of the inhomogeneities effects on the dynamics of the
universe and the deviations that can possibly be introduced with
respect to the simple FRW models
\cite{thomas,syksy,wiltshire,pedro,roy}.

\acknowledgments I thank Yuri V. Baryshev and Daniil Tekhanovich for
useful discussions and comments.  I acknowledge the use of the Sloan
Digital Sky Survey data ({\tt http://www.sdss.org}), of the NYU
Value-Added Galaxy Catalog ({\tt http://ssds.physics.nyu.edu/}) and of
the Millennium run semi-analytic galaxy catalog ({\tt
  http://www.mpa-garching.mpg.de/galform/agnpaper}).


\begin{thebibliography}{99}   


\bibitem{zehavi} Zehavi, I., et al., Astrophys.J., {\bf 571}, 172
  (2002)


\bibitem{Hawkins} Hawkins, E. et al., Mon.Not.R.Acad.Soc., {\bf 346},
  77 (2003)

\bibitem{dr4_paper} Sylos Labini, F., Vasilyev, N.L.  Baryshev, Yu.V.,
  Astron.Astrophys., {\bf 465}, 23 (2007)

\bibitem{shanks1} Frith, W.J. et al.,
Mon.Not.R.Acad.Soc., {\bf 345}, 1049 (2003)

\bibitem{shanks2} Busswell G. S. et al.,  
Mon.Not.R.Acad.Soc.  {\bf 354} 991 (2004)

\bibitem{shanks3} Frith, W. J., et al. 
Mon.Not.R.Acad.Soc., {\bf 371}, 1601 (2006) 

\bibitem{einasto} Einasto J., et al., Astron.Astrophys., {\bf 459}, 1
  (2006)

\bibitem{2df_epl} Sylos Labini F.  Vasilyev, N.L.  Baryshev, Yu.V.,
  Europhys.Lett. , {\bf 85 } 29002 (2009)


\bibitem{2df_aea} 
Sylos Labini F.  Vasilyev, N.L.  Baryshev, Yu.V.,
Astron.Astrophys  {\bf 496}  7 (2009)  

\bibitem {saslaw} Yang A. and Saslaw
  W.C. Astrophys.J.  {\bf 729} 1 (2011)


\bibitem{einasto11} Einasto J., et al., Astrophys.J., in print (2011)

\bibitem {pee93} Peebles, P.J.E., {\it ``Principles Of Physical
  Cosmology''}, (Princeton University Press, Princeton, New Jersey,
  1993)

\bibitem{glass} Gabrielli, A., et al.
Physical Review {\bf D65}, 083523 (2002)

\bibitem{martinez_bao} Mart\'inez, V.J., et
  al., Astrophys.J., {\bf 696}, L93 (2009)

\bibitem{bao_aea} Sylos Labini F., et al.,  
2009  Astron.Astrophys.,  {\bf 505} 981

\bibitem{kazin} Kazin et al., 
Astrophys.J., {\bf 710}, 1444 (2010)

\bibitem{lahav11} Thomas S.A., Abdalla, F.B., Lahav O., 
  Phys. Rev. Lett. {\bf 106}, 241301 (2011)

\bibitem{shanks4} Sawangwit U. et al., Mon.Not.R.Acad.Soc. in the
  press (2011) 

\bibitem{york}  York, D., et al.,
 Astron.J.,  {\bf 120}, 1579 (2000)

\bibitem{dr6} Adelman-McCarthy,
  J.K., et al., Astrophys.J.Suppl.. {\bf 175}, 297 (2008)

\bibitem{hogg}
Hogg D.W.,  et al.,Astrophys.J., {\bf 624}, 54 (2005)

\bibitem{tibor} Antal, T., Sylos Labini, F.,
  Vasilyev, N.L., Baryshev, Yu. V., Europhys.Lett. {\bf 88}, 59001
  (2009)
\bibitem{sdss_epl}  Sylos Labini F., et al., 
  Europhysics Letters, {\bf 86} 49001 (2009)  
  
\bibitem{sdss_aea} Sylos Labini, F., Vasilyev, N.L., Baryshev, Yu. V.,
  {\bf 508}, 17 (2009)

\bibitem{loveday}Loveday, J.,  2004,
Mon.Not.R.Acad.Soc.  {\bf 347} 601 (2004)

\bibitem{blanton2003} 
Blanton, M.R., et al.
 Astrophys.J. {\bf 592}, 819 (2003)


\bibitem{dr7} Abazajian K., et al., 
  Astrophys.J.Suppl. {\bf 182}  543 (2009)

\bibitem{strauss} Strauss, M.A., et al.,
 Astron.J., {\bf  124}, 1810 (2002)

\bibitem{vagc} Blanton M. R. et al. 
 Astron.J.,  {\bf  129} 2562 (2005)

\bibitem{book} Gabrielli A., Sylos Labini F.,
  Joyce M., Pietronero L., 2005, {\it Statistical Physics for Cosmic
    Structures} (Springer Verlag, Berlin)

\bibitem{copernican} Sylos Labini F.,  Baryshev Y. V. 
{\bf JCAP06} (2010) 021 

\bibitem{hog} Hog E. et al., 
Astron. Astrophys. {\bf 355}, L27  (2000)

\bibitem {springel} Springel, V., et
  al.,Nature, {\bf 435}, 629 (2005)

\bibitem{croton} Croton, D.J. et al., 
Mon.Not.R.Acad.Soc., {\bf 365}, 11 (2006)



\bibitem {cdm_theo} Sylos Labini, F. and Vasilyev, N.L.,
  Astron.Astrophys. {\bf 477}, 381 (2008)


\bibitem{syksy}  R\"{a}s\"{a}nen S. 
Int.J.Mod.Phys., {\bf D17} 2543 (2008)


\bibitem{thomas} Buchert T. 
Gen.Rel.Grav., {\bf 40} 467 (2008) 

\bibitem{wiltshire} Wiltshire D. L.,
  Int.J.Mod.Phys., {\bf 17},
  641 (2008) 

\bibitem{pedro} Clifton T., Ferreira, P. G., Phys.Rev. {\bf D80}
  103503 (2009)
	
\bibitem{roy} Maartens R.  {\tt arXiv:1104.1300 } (2011)
 
\end{thebibliography}
\end{document}